\documentclass[pra,twocolumn,showpacs]{revtex4}
\usepackage{graphicx}

\begin{document}
\title{Production of genuine entangled states of four atomic qubits}
\author{Gui-Yun Liu and  Le-Man Kuang\footnote{Author to whom any correspondence should be
addressed. }\footnote{ Email: lmkuang@hunnu.edu.cn}}
\address{Key Laboratory of Low-Dimensional Quantum Structures and Quantum Control of Ministry of Education,
and Department of Physics, Hunan Normal University, Changsha 410081,
China}

\begin{abstract}
We propose an optical scheme to generate genuine entangled states
of four atomic qubits in optical cavities using a single-photon
source, beam splitters and single photon detectors. We show how to
generate deterministically sixteen orthonormal and independent
genuine entangled states of four atomic qubits. It is found that
the sixteen genuine entangled states form a new type of
representation of the four-atomic-qubit system, i.e., the genuine
entangled-state representation. This representation brings new
interesting insight onto better understanding multipartite
entanglement.
\end{abstract}

\maketitle

\section{Introduction}

Quantum entanglement plays an important role in quantum
information processing and quantum mechanics. While bipartite
entanglement is well understood, multiparty entanglement is still
under intensive research. Multipartite entanglement is thus not a
straightforward extension of bipartite entanglement and gives rise
to new phenomena which can be exploited in quantum information and
quantum computing processes. For example, there are quantum
communication protocols that require multiparty entanglement such
as universal error correction \cite{shor}, quantum secret sharing
\cite{hill}, and telecloning \cite{mura}. Also, highly entangled
multipartite states are needed for one-way quantum computing
\cite{raus}. In fact, all known quantum algorithms do work with
multipartite entanglement. On the other hand, multipartite
entangled states provide a stronger test of local realism. As a
general rule, one can say that the more particles are entangled,
the more clearly nonclassical effects are exhibited, and the more
useful the states are for quantum applications. In addition,
multipartite entanglement is expected to play a key role on
quantum-phase transition  phenomena \cite{oliv}.

Although multipartite entanglement is ubiquitous in many-body
quantum systems, it is very difficult both to characterize and to
quantify it. Up to now, there is no unique way to define
multipartite entanglement, even in the simplest case of pure
states. The presence of multipartite entanglement clearly depends
on the partitioning that one imposes in order to group the
individual subsystems into parties. Furthermore, given a fixed
partition, one can single out a hierarchy of different levels of
multipartite entanglement which establishes a smooth connection
between the two limiting cases of a fully separable state, where
the parties are all disentangled, and of fully inseparable states,
where entanglement exists across any global bisection, and the
parties are supposed to share genuine multipartite entanglement.
Genuine multipartite entanglement is distinguished from other
types of entanglement by the participation of all parties in
quantum correlations, and it is particularly distinct from
biseparable entanglement. Yeo and Chua \cite{yeo} indicated that
an arbitrary two-qubit state can be faithfully teleported by the
use of a genuine four-qubit entangled state. And it was found that
a genuine $2N$-qubit entangled state can be used to teleport an
arbitrary $N$-qubit state \cite{chen}, and a genuine
$(2N+1)$-qubit entangled state can realize controlled
teleportation of an arbitrary $N$-qubit state \cite{man}.  One of
the important issues regarding many-body quantum systems is to
generate and to verify genuine multipartite entanglement among
parties. The purpose of this paper is to propose an optical scheme
to produce genuine entangled states (GESs) of four atomic qubits
in optical cavities. This paper is organized as follows. In
section 2, we propose our theoretical model. In section 3, we show
how to create sixteen orthonormal and independent genuine
entangled states of four atomic qubits. We shall conclude our
paper with discussions and remarks in the last section.

\section{Theoretical model}

The basic setup for genuine entanglement generation of four atomic
qubits is indicated in Fig. 1 where we make use a Mach-Zehnder
(MZ) interferometer consisting of two 50/50 optical beam splitters
(BSs). The input light field is bifurcated at the first BS, guided
to interact sequentially with the atomic qubits which are placed
in four separate high-finesse optical cavities in which four
atomic qubit are placed, and then recombined at the second BS.  We
consider atomic qubits based on degenerate hyperfine states
denoted by $|0\rangle_i$ and $|1\rangle_i$ ($i=1,2,3,4$). A single
pulse of light is passed through an optical interferometer, with
the different ¡°arms¡± of the interferometer corresponding to
different photon polarization states. Each polarization state
interacts with a different internal atomic state through the
mechanism proposed in Ref. \cite{dua}. Under the condition of the
large detuning, the atom-photon evolution can be described by the
following unitary transformation \cite{hua}
\begin{eqnarray}
\label{1}
\hat{U}_{i}=\exp[-i\phi(\hat{\emph{a}}^{\dagger}_{U}\hat{\emph{a}}_{U}|0\rangle_{i}\langle0|
+\hat{\emph{a}}^{\dagger}_{L}\hat{\emph{a}}_{L}|1\rangle_{i}\langle1|)],
\end{eqnarray}
where $\hat{\emph{a}}^{\dagger}_{U,L}$ and $\hat{\emph{a}}_{U,L}$
are the creation and annihilation operators of the optical fields
corresponding to the upper arm and the low arm, respectively.
$|0\rangle_{i}$ and $|1\rangle_{i}$ are the atomic degenerate
hyperfine states corresponding to the four atomic qubits
($i=1,2,3,4$). The interaction is governed by the phase-shift $
\phi=\frac{\textit{d}^{2}\mathcal {E}^{2}\tau}{\hbar^{2}\Delta}$,
where $\textit{d}$ is the electrical dipole moment and $\mathcal
{E}=\sqrt{{\hbar\omega}/{2\epsilon_{0}V}}$ is the electrical field
of a photon for laser with the frequency $\omega$ and the mode
volume $\textit{V}$, $\tau$ is the atom-photon interaction time,
$\Delta$ is the detuning between the atomic resonance and laser
frequencies.  The unitary evolution of the whole system indicated
in Fig. 1 can be described by the following unitary transformation
\begin{eqnarray}
\label{2}
\hat{U}=\hat{U}_{BS2}\hat{U}_{4}\hat{U}_{3}\hat{U}_{2}\hat{U}_{1}\hat{U}_{BS1},
\end{eqnarray}
where $\hat{U}_{BSi}$ ($i=1,2$)is the 50/50 BS operator given by
\begin{eqnarray}
\label{3}
\hat{U}_{BSi}=\exp\left[-i\frac{\pi}{4}(\hat{\emph{a}}^{\dagger}_{U}\hat{\emph{a}}_{L}+\hat{\emph{a}}^{\dagger}_{L}\hat{\emph{a}}_{U})\right].
\end{eqnarray}

In the following we will show that genuine entangled states of
four atomic qubits by using above unitary transformation and
single photon detections.

\begin{figure}[htp]
\begin{center}
\includegraphics[width=8.3cm,height=2cm]{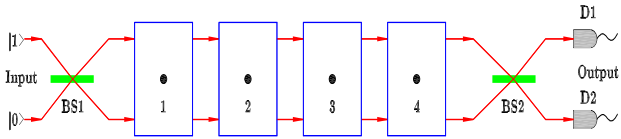}
\caption{Schematic setup to generate genuine four atomic qubits.
The input single-photon field is bifurcated at the first BS,
guided to interact sequentially with the atomic qubits placed in
four separate high-finesse optical cavities, and then recombined
at the second BS. Each cavity has an atomic qubit. D1 and D2 are
two single photon detectors.}
\end{center}
\end{figure}

\section{Genuine entangled states of four atomic qubits}

In this section, we show how to prepare genuine entangled states
of four atomic qubits  using the setup indicated in Fig. 1. Let us
consider the case of the single-photon input. In this case, the
upper channel is in the single-photon state $|1\rangle$ while the
lower channel is in the vacuum state $|0\rangle$. Then the initial
state of the optical field is $|\Psi_{i}\rangle_{O}=|10\rangle$
while the initial state of the four atomic qubits can be supposed
as
\begin{eqnarray}
\label{4}
|\Psi_{i}\rangle_{A}=|\Phi_{1}\rangle\otimes|\Phi_{2}\rangle\otimes|\Phi_{3}\rangle\otimes|\Phi_{4}\rangle,
\end{eqnarray}
where the initial state of the single atomic qubit is given  by
\begin{equation}
\label{5} |\Phi_{i}\rangle=\cos\theta_{i}|0\rangle_{i} +
\sin\theta_{i}|1\rangle_{i}, \hspace{0.3cm}(i=1,2,3,4),
\end{equation}
which implies that the initial state of the whole system is $
|\Psi_{i}\rangle=|\Psi_{i}\rangle_{O}\otimes|\Psi_{i}\rangle_{A}$.

The state of the system at the output of the MZ interferometer is
then given by $|\Psi_{f}\rangle=\hat{U}|\Psi_{i}\rangle$ where the
unitary transformation $\hat{U}$ is given by Eq. (3).  It is easy
to find that the output state $|\Psi_{f}\rangle$ has the following
expression
\begin{eqnarray}
\label{6}
|\Psi_{f}\rangle=|01\rangle\otimes|\chi^{'}(\phi)\rangle+|10\rangle\otimes|\chi^{''}(\phi)\rangle,
\end{eqnarray}
where we have got rid of a global phase factor,  $|01\rangle$ and
$|10\rangle$ are quantum states of two light fields,
$|\chi^{''}(\phi)\rangle$ and $|\chi^{''}(\phi)\rangle$ are
quantum states of the four atomic qubits given by
\begin{eqnarray}
\label{7-8} \left\vert\chi^{^{\prime}}(\phi)\right\rangle&=&
\cos2\phi
|A_{+}\rangle+\cos\phi (|B\rangle+|C\rangle)+|D\rangle,\\
\left\vert\chi^{^{\prime \prime }}(\phi)\right\rangle&=&\sin2\phi
A_{-}\rangle +\sin\phi(|B \rangle-|C\rangle),
\end{eqnarray}
which  are generally entangled states. Here we have introduced
\begin{eqnarray}
\label{9-12} \left\vert A_{\pm
}\right\rangle&=&c_{1}c_{2}c_{3}c_{4}\left\vert 0000\right\rangle
\pm
s_{1}s_{2}s_{3}s_{4}\left\vert 1111\right\rangle,  \\
\left\vert B\right\rangle&=&c_{1}c_{2}s_{3}c_{4}\left\vert
0010\right\rangle + c_{1}s_{2}c_{3}c_{4}\left\vert
0100\right\rangle \nonumber
\\&& + s_{1}c_{2}c_{3}c_{4}\left\vert
1000\right\rangle + c_{1}c_{2}c_{3}s_{4}\left\vert 0001\right\rangle,  \\
\left\vert C\right\rangle &=&s_{1}s_{2}s_{3}c_{4}\left\vert
1110\right\rangle + c_{1}s_{2}s_{3}s_{4}\left\vert
0111\right\rangle \nonumber\\&&+ s_{1}s_{2}c_{3}s_{4}\left\vert
1101\right\rangle + s_{1}c_{2}s_{3}s_{4}\left\vert
1011\right\rangle],  \\
\left\vert D\right\rangle  &=&c_{1}s_{2}s_{3}c_{4}\left\vert
0110\right\rangle + s_{1}s_{2}c_{3}c_{4}\left\vert
1100\right\rangle \nonumber\\
&&+ s_{1}c_{2}s_{3}c_{4}\left\vert 1010\right\rangle + c_{1}c_{2}s_{3}s_{4}\left\vert 0011\right\rangle\nonumber\\
&& + c_{1}s_{2}c_{3}s_{4}\left\vert 0101\right\rangle _{a}+
s_{1}c_{2}c_{3}s_{4}\left\vert 1001\right\rangle,
\end{eqnarray}
where we have used notations $c_i=\cos\theta_{i}$ and
$s_i=\sin\theta_{i}$.

Eq. (6) indicates that entangled states of the four atomic qubits
can be produced through making quantum measurements upon quantum
states of the output light and controlling the phase $\phi$, i.e.,
the photon-atom interaction time  $\tau$. In fact, when
$\phi=\pi/2$, if the photodetector D2 is triggered, at the same
time the photodetector D1 detects the null result, the
atomic-qubit state will collapse onto the following superposition
state
\begin{eqnarray}
|{\chi}^{'}(\pi/2)\rangle&=&\frac{1}{\sqrt{\Gamma_{1}}}(|\Lambda^{'}_1\rangle
+ |\Lambda^{'}_2\rangle),
\end{eqnarray}
where $|\Lambda^{'}_1\rangle$ and $|\Lambda^{'}_2\rangle$ are
defined by
\begin{eqnarray}
|\Lambda^{'}_1\rangle&=&c_{1}c_{2}c_{3}c_{4}|0000\rangle- c_{1}s_{2}s_{3}c_{4}|0110\rangle\nonumber\\
&&- c_{1}c_{2}s_{3}s_{4}|0011\rangle - c_{1}s_{2}c_{3}s_{4}|0101\rangle,\\
|\Lambda^{'}_2\rangle&=&  s_{1}s_{2}s_{3}s_{4}|1111\rangle-
s_{1}s_{2}c_{3}c_{4}|1100\rangle\nonumber\\
&&- s_{1}c_{2}s_{3}c_{4}|1010\rangle
-s_{1}c_{2}c_{3}s_{4}|1001\rangle),
\end{eqnarray}
where the normalization constant is given by
\begin{eqnarray}
\Gamma_{1}=\frac{1}{2}\left( 1+\cos 2\theta _{1}\cos 2\theta
_{2}\cos 2\theta _{3}\cos 2\theta _{4}\right).
\end{eqnarray}

Similarly, when $\phi=\pi/2$, if the photodetector D1 is
triggered, and the photodetector D2 detects the null result, the
atomic qubits state will become the following superposition state
\begin{equation}
|{\chi}^{''}(\pi/2)\rangle=\frac{1}{\sqrt{\Gamma_{1}}}(|\Lambda^{''}_1\rangle
- |\Lambda^{''}_2\rangle)
\end{equation}
where  $|\Lambda^{''}_1\rangle$ and $|\Lambda^{''}_2\rangle$ are
defined by
\begin{eqnarray}
|\Lambda^{''}_1\rangle&=& s_1s_2s_3c_4|1110\rangle +
c_1s_2s_3s_4|0111\rangle\nonumber\\&& + s_1s_2c_3s_4|1101\rangle
+ s_1c_2s_3s_4|1011\rangle, \\
|\Lambda^{''}_2\rangle&=& c_1c_2s_3c_4|0010\rangle+ c_1s_2c_3c_4|0100\rangle\nonumber\\
&&+ s_1c_2c_3c_4|1000\rangle + c_1c_2c_3s_4|0001\rangle),
\end{eqnarray}
where the normalization constant is given by
\begin{eqnarray}
\Gamma_{2}=\frac{1}{2}\left( 1-\cos 2\theta _{1}\cos 2\theta
_{2}\cos 2\theta _{3}\cos 2\theta _{4}\right).
\end{eqnarray}

It is obvious to see that quantum states of the four atomic qubits
$|{\chi}^{'}(\pi/2)\rangle$ and $|{\chi}^{''}(\pi/2)\rangle$ given
by Eqs. (13) and (17) are generally  entangled states. In what
follows we show that under certain conditions,
$|{\chi}^{'}(\pi/2)\rangle$ and $|{\chi}^{''}(\pi/2)\rangle$ will
be GESs of four atomic qubits. In order to observe this, we now
investigate entanglement properties of entangled states
$|{\chi}^{'}(\pi/2)\rangle$ and $|{\chi}^{''}(\pi/2)\rangle$ in
which quantum state of any two atomic qubits is a mixed state. The
degree of entanglement of a two-qubit mixed state can be measured
in terms of the concurrence function defined in Ref. \cite{woo}.
It is straightforward to find the concurrence between any two
atomic qubits for quantum states $|{\chi}^{'}(\pi/2)\rangle$ and
$|{\chi}^{''}(\pi/2)\rangle$ to be
\begin{eqnarray}
C(|\chi^{'}\rangle)&=&\max \left(0,\lambda_{+}\right),
C(|\chi^{''}\rangle)=\max \left(0,\lambda_{-}\right).
\end{eqnarray}
where the two parameters $\lambda_{\pm}$ are given by
\begin{equation}
\lambda_{\pm}=\frac{\left\vert \cos 2\theta _{1}\cos
2\theta_{2}\sin 2\theta_{3}\sin 2\theta_{4}\right\vert }{1\pm \cos
2\theta_{1}\cos 2\theta_{2}\cos 2\theta_{3}\cos 2\theta_{4}}.
\end{equation}

The amount of entanglement between any two atom-qubit pairs in
four-atom-qubit entangled states $|{\chi}^{'}(\pi/2)\rangle$ and
$|{\chi}^{''}(\pi/2)\rangle$ can be measured in terms of the von
Neumann entropy \cite{yeo}. We find that the von Neumann entropy
for any two qubit-pairs in four-atom-qubit entangled states
$|{\chi}^{'}(\pi/2)\rangle$ and $|{\chi}^{''}(\pi/2)\rangle$ is
given by
\begin{eqnarray}
S(|\chi^{'}\rangle)&=&1-\frac{1}{2}[(1+\delta _{+})\log
_{2}(1+\delta _{+})\nonumber\\
&&+(1-\delta_{+})\log _{2}(1-\delta _{+})],\\
S(|\chi^{''}\rangle)&=&1-\frac{1}{2}[(1+\delta_{-})\log _{2}(1+\delta _{-})\nonumber\\
&&+(1-\delta _{-})\log _{2}(1-\delta _{-})],
\end{eqnarray}
where we have introduced
\begin{eqnarray}
\delta _{\pm} &=&\frac{\cos 2\theta _{3}\cos 2\theta _{4}\pm\cos
2\theta _{1}\cos 2\theta _{2}}{1\pm\cos 2\theta _{1}\cos 2\theta
_{2}\cos 2\theta _{3}\cos 2\theta _{4}}.
\end{eqnarray}

A further calculation indicates that the amount of entanglement
between any one atomic qubit and the other three atomic qubits in
four-atom-qubit entangled states $|{\chi}^{'}(\pi/2)\rangle$ and
$|{\chi}^{''}(\pi/2)\rangle$ is the same as that of any two
atomic-qubit pairs.

In particular, from Eqs. (13) and (17) we can see that when
$\theta_{1}=\theta_{1}=\theta_{2}=\theta
_{3}=\theta_{4}=\pm\pi/4$, the resultant entangled states become
\begin{eqnarray}
|\overline{{\chi}^{'}}\rangle&=&\frac{1}{2\sqrt{2}}(|0000\rangle+|1111\rangle-|0110\rangle-|1100\rangle\nonumber\\
&&-|1010\rangle-|0011\rangle-|0101\rangle-|1001\rangle),\\
|\overline{{\chi}^{''}}\rangle&=&\frac{1}{2\sqrt{2}}(|1110\rangle+|0111\rangle+|1101\rangle+|1011\rangle\nonumber\\
&&-|0010\rangle-|0100\rangle-|1000\rangle-|0001\rangle).
\end{eqnarray}

For above entangled states, making Eqs. (21-25) we find that
$C(|\chi^{'}\rangle)=C(|\chi^{''}\rangle)=0$ and
$S(|\chi^{'}\rangle)=S(|\chi^{''}\rangle)=1$. This implies that
there is absolutely zero entanglement between nay one atomic qubit
and any other atomic qubit, and the entanglement is purely between
pairs of atomic qubits. Therefore, these entangled states given by
Eqs. (26) and (27) are GESs for four atomic qubits.

We then have a look at the probability of success to get above
GESs. From Eq. (6), it is straightforward to see that when the
second photodetector D2 clicks, at the same time the photodetector
D1 detects the null result, the atomic-qubit state will collapse
onto the GES $|\overline{{\chi}^{'}}\rangle$ with the probability
of success being $1/2$, and  when the first photodetector D1
clicks, at the same time the photodetector D2 detects the null
result, the atomic-qubit state will collapse onto the GES
$|\overline{{\chi}^{''}}\rangle$ with the probability of success
being $1/2$ too.  It is interesting to note that the GES
$|\overline{{\chi}^{''}}\rangle$ (or
$|\overline{{\chi}^{'}}\rangle$) can be obtained by applying the
Pauli operator $\hat{\sigma}^y_4$ to fourth atomic qubit of the
GES $|\overline{{\chi}^{''}}\rangle$ (or
$|\overline{{\chi}^{'}}\rangle$), i.e.,
\begin{equation}
\hat{\sigma}^y_4|\overline{{\chi}^{'}}\rangle=
|\overline{{\chi}^{''}}\rangle, \hspace{0.3cm}
\hat{\sigma}^y_4|\overline{{\chi}^{''}}\rangle=
|\overline{{\chi}^{'}}\rangle,
\end{equation}
which implies that the GESs $|\overline{{\chi}^{'}}\rangle$ and
$|\overline{{\chi}^{''}}\rangle$  can be produced with the
probability of success being one through making use of
$\hat{\sigma}^y$ operation on fourth atomic qubit. Hence, in our
scheme the GESs $|\overline{{\chi}^{'}}\rangle$ and
$|\overline{{\chi}^{''}}\rangle$  can be produced
deterministically through making single photon detections of
output fields and unitary transformation ($\hat{\sigma}^y$) upon
the fourth atomic qubit.

It is interesting to see that starting with the GES
$|\overline{{\chi}^{'}}\rangle$ or
$|\overline{{\chi}^{''}}\rangle$ we can generate a basis of
sixteen orthonormal states by applying Pauli operators of atomic
qubits to the GES $|\overline{{\chi}^{'}}\rangle$ or
$|\overline{{\chi}^{''}}\rangle$. For instance, for the GES
$|\overline{{\chi}^{'}}\rangle$ we can obtain the following
sixteen GESs
\begin{eqnarray}
|\varphi_1\rangle_\mu&=&\sigma _{1}^{0}\sigma
_{2}^{\mu}|\overline{{\chi}^{'}}\rangle, \hspace{0.3cm}
|\varphi_2\rangle_\mu=\sigma _{1}^{3}\sigma
_{2}^{\mu}|\overline{{\chi}^{'}}\rangle,   \\
|\varphi_3\rangle_\mu&=&\sigma _{1}^{0}\sigma _{2}^{\mu}\sigma
_{3}^{3}|\overline{{\chi}^{'}}\rangle, \hspace{0.3cm}
|\varphi_4\rangle_\mu=\sigma _{1}^{3}\sigma _{2}^{\mu}\sigma
_{3}^{3}|\overline{{\chi}^{'}}\rangle,
\end{eqnarray}
where $\sigma _{i}^{\mu}$ ($\mu=0,1,2,3$) denotes the $\mu$-th
component of the Pauli operator for the $i$-th atomic qubit with
$\sigma _{i}^{0}$ being the unit matrix. The above sixteen GESs
can be explicitly expressed as
\begin{eqnarray}
|\varphi_1\rangle_0&=&\frac{1}{\sqrt{8}}( \left\vert
0000\right\rangle +\left\vert 1111\right\rangle -\left\vert
0110\right\rangle -\left\vert 1100\right\rangle
\nonumber\\&&-\left\vert 1010\right\rangle -\left\vert
0011\right\rangle -\left\vert 0101\right\rangle -\left\vert
1001\right\rangle ), \\
|\varphi_1\rangle_1&=&\frac{1}{\sqrt{8}}( -\left\vert
1110\right\rangle -\left\vert 0111\right\rangle -\left\vert
1101\right\rangle +\left\vert 1011\right\rangle \nonumber\\&&
-\left\vert 0010\right\rangle +\left\vert 0100\right\rangle
-\left\vert 1000\right\rangle -\left\vert
0001\right\rangle), \\
|\varphi_1\rangle_2&=&\frac{1}{\sqrt{8}}( \left\vert
1110\right\rangle +\left\vert 0111\right\rangle +\left\vert
1101\right\rangle +\left\vert 1011\right\rangle \nonumber\\&&
-\left\vert 0010\right\rangle -\left\vert 0100\right\rangle
-\left\vert 1000\right\rangle -\left\vert
0001\right\rangle ), \\
|\varphi_1\rangle_3&=&\frac{1}{\sqrt{8}}( -\left\vert
0000\right\rangle +\left\vert 1111\right\rangle -\left\vert
0110\right\rangle -\left\vert 1100\right\rangle \nonumber\\&&
+\left\vert 1010\right\rangle +\left\vert 0011\right\rangle
-\left\vert 0101\right\rangle +\left\vert
1001\right\rangle), \\
|\varphi_2\rangle_0&=&\frac{1}{\sqrt{8}}( -\left\vert
0000\right\rangle +\left\vert 1111\right\rangle +\left\vert
0110\right\rangle -\left\vert 1100\right\rangle
\nonumber\\&&
-\left\vert 1010\right\rangle +\left\vert
0011\right\rangle +\left\vert 0101\right\rangle -\left\vert
1001\right\rangle, \\
|\varphi_2\rangle_1&=&\frac{1}{\sqrt{8}}( -\left\vert
1110\right\rangle +\left\vert 0111\right\rangle -\left\vert
1101\right\rangle +\left\vert 1011\right\rangle
\nonumber\\&&
+\left\vert 0010\right\rangle -\left\vert
0100\right\rangle -\left\vert 1000\right\rangle +\left\vert
0001\right\rangle), \\
|\varphi_2\rangle_2&=&\frac{1}{\sqrt{8}}( \left\vert
1110\right\rangle -\left\vert 0111\right\rangle +\left\vert
1101\right\rangle +\left\vert 1011\right\rangle
\nonumber\\&&
+\left\vert 0010\right\rangle +\left\vert
0100\right\rangle -\left\vert
1000\right\rangle +\left\vert 0001\right\rangle), \\
|\varphi_2\rangle_3&=&\frac{1}{\sqrt{8}}(\left\vert
0000\right\rangle +\left\vert 1111\right\rangle +\left\vert
0110\right\rangle -\left\vert 1100\right\rangle \nonumber\\&&
+\left\vert 1010\right\rangle -\left\vert 0011\right\rangle
+\left\vert 0101\right\rangle +\left\vert 1001\right\rangle),\\
|\varphi_3\rangle_0&=&\frac{1}{\sqrt{8}}( -\left\vert
0000\right\rangle +\left\vert 1111\right\rangle -\left\vert
0110\right\rangle +\left\vert 1100\right\rangle \nonumber\\&&
-\left\vert 1010\right\rangle -\left\vert 0011\right\rangle
+\left\vert 0101\right\rangle +\left\vert
1001\right\rangle), \\
|\varphi_3\rangle_1&=&\frac{1}{\sqrt{8}}( -\left\vert
1110\right\rangle -\left\vert 0111\right\rangle +\left\vert
1101\right\rangle +\left\vert 1011\right\rangle
\nonumber\\&&
-\left\vert 0010\right\rangle -\left\vert
0100\right\rangle +\left\vert 1000\right\rangle +\left\vert
0001\right\rangle), \\
|\varphi_3\rangle_2&=&\frac{1}{\sqrt{8}}( \left\vert
1110\right\rangle +\left\vert 0111\right\rangle -\left\vert
1101\right\rangle +\left\vert 1011\right\rangle
\nonumber\\&&
-\left\vert 0010\right\rangle +\left\vert
0100\right\rangle +\left\vert 1000\right\rangle +\left\vert
0001\right\rangle), \\
|\varphi_3\rangle_3&=&\frac{1}{\sqrt{8}}( \left\vert
0000\right\rangle +\left\vert 1111\right\rangle -\left\vert
0110\right\rangle +\left\vert 1100\right\rangle
\nonumber\\&&
+\left\vert 1010\right\rangle +\left\vert
0011\right\rangle +\left\vert 0101\right\rangle -\left\vert
1001\right\rangle), \\
|\varphi_4\rangle_0&=&\frac{1}{\sqrt{8}}( \left\vert
0000\right\rangle +\left\vert 1111\right\rangle +\left\vert
0110\right\rangle +\left\vert 1100\right\rangle
\nonumber\\&&
-\left\vert 1010\right\rangle +\left\vert
0011\right\rangle -\left\vert 0101\right\rangle +\left\vert
1001\right\rangle), \\
|\varphi_4\rangle_1&=&\frac{1}{\sqrt{8}}( -\left\vert
1110\right\rangle +\left\vert 0111\right\rangle +\left\vert
1101\right\rangle +\left\vert 1011\right\rangle
\nonumber\\&&
+\left\vert 0010\right\rangle +\left\vert
0100\right\rangle +\left\vert 1000\right\rangle -\left\vert
0001\right\rangle), \\
|\varphi_4\rangle_2&=&\frac{1}{\sqrt{8}}( \left\vert
1110\right\rangle -\left\vert 0111\right\rangle -\left\vert
1101\right\rangle +\left\vert 1011\right\rangle
\nonumber\\&&
+\left\vert 0010\right\rangle -\left\vert
0100\right\rangle +\left\vert
1000\right\rangle -\left\vert 0001\right\rangle),\\
|\varphi_4\rangle_3&=&\frac{1}{\sqrt{8}}( -\left\vert
0000\right\rangle +\left\vert 1111\right\rangle +\left\vert
0110\right\rangle +\left\vert 1100\right\rangle \nonumber\\&&
+\left\vert 1010\right\rangle -\left\vert 0011\right\rangle
-\left\vert 0101\right\rangle -\left\vert 1001\right\rangle).
\end{eqnarray}

It is straightforward to check that
$_\nu\langle\varphi_\mu|\varphi_{\mu'}\rangle_{\upsilon'}=\delta_{\mu\mu'}\delta_{\nu\nu'}$
and
$\sum_{\mu\nu}|\varphi_\mu\rangle_{\nu\nu}\langle\varphi_\mu|=1$.
This implies that above sixteen states form a orthonormal and
completeness Hilbert space of the four-qubit system. In fact, they
build a new type representation of the four-qubit system, i.e., a
genuine entangled-state representation. In this representation an
arbitrary state of the four-qubit system can be expressed in terms
of the basis of the representation. In order to see this, we
consider the typical four-qubit entangled states
\cite{gre,dur,bri,dic}: the GHZ state $|\tt{GHZ}_4\rangle$, the
$W$ state $|W_4\rangle$, the cluster state $\left\vert
\tt{CL}_4\right\rangle$, and the symmetric Dicke state
$|D_4\rangle$. They have the following expressions, respectively,
\begin{eqnarray}
|\tt{GHZ}_4\rangle&=&\frac{1}{\sqrt2}(|0000\rangle +
|1111\rangle), \\
|W_4\rangle&=&\frac{1}{2}(|0001\rangle+|0010\rangle+|0100\rangle+|1000\rangle), \\
\left\vert \tt{CL}_4\right\rangle&=&\frac{1}{2}( \left\vert
0000\right\rangle +\left\vert 0110\right\rangle +\left\vert
1001\right\rangle -\left\vert 1111\right\rangle ), \\
|D_4\rangle&=&\frac{1}{\sqrt6}(|0011\rangle
+|0101\rangle+|1001\rangle\nonumber\\
&&+|1100\rangle+|0110\rangle+|1010\rangle).
\end{eqnarray}

Making use of the basis of the genuine entangled-state
representation given by Eqs. (31-46), we can find that above four
typical four-qubit entangled states can be expressed as follows:
\begin{eqnarray}
|\tt{GHZ}_4\rangle&=&\frac{1}{2}\left(\left\vert \varphi
_{1}\right\rangle_{0}+\left\vert\varphi_{3}\right\rangle
_{3}+\left\vert \varphi_{2}\right\rangle_{3}+\left\vert \varphi
_{4}\right\rangle
_{0}\right),  \\
\left\vert W_4\right\rangle&=&\frac{1}{\sqrt{8}}(\left\vert
\varphi_{3}\right\rangle_{2}+\left\vert \varphi_{2}\right\rangle
_{2}-\left\vert
\varphi_{1}\right\rangle_{1}-2\left\vert\varphi_{1}\right\rangle_{2}\nonumber
\\&&+\left\vert
\varphi _{4}\right\rangle _{1}),\\
\left\vert \tt{CL}_4\right\rangle&=&\frac{1}{\sqrt{8}}( \left\vert
\varphi _{2}\right\rangle _{3}+\left\vert \varphi
_{4}\right\rangle
_{0}-\left\vert \varphi _{1}\right\rangle _{0} -\left\vert\varphi_{3}\right\rangle _{0}\nonumber\\
&&-\left\vert \varphi_{1}\right\rangle _{3}-\left\vert \varphi
_{3}\right\rangle _{3}-\left\vert \varphi _{4}\right\rangle
_{3}-\left\vert \varphi _{2}\right\rangle _{0}),\\
\left\vert D_{4}\right\rangle &=&\frac{1}{2\sqrt{3}}( 2\left\vert
\varphi _{2}\right\rangle _{3}-\left\vert \varphi
_{4}\right\rangle _{3}+\left\vert \varphi_{2}\right\rangle_{0}
+\left\vert\varphi _{1}\right\rangle _{3}\nonumber\\&&-\left\vert
\varphi _{3}\right\rangle _{0}-2\left\vert \varphi
_{1}\right\rangle_{0}).
\end{eqnarray}

Finally, we consider the influence of the imperfection of photon
detections in the present scheme. An ideal photon detection with
quantum efficiency $\eta=1$ can be described by the positive
operator-valued measure (POVM) of each detector
$\{\Pi_0=|0\rangle\langle0|,\Pi_1=I-\Pi_0\}$. In the realistic
case, an incoming photon can not be detected with the probability
success 1. If the quantum efficiency of the photodetector is
$\eta$, the POVM is given by \cite{kok,kua,oli}
\begin{equation}
\label{31} \Pi_0(\eta)=\sum_{i=0}(1-\eta)^{i}|i\rangle\langle
i|,~~~\Pi_1(\eta)=I-\Pi_0(\eta),
\end{equation}
from which it is straightforward to find that the inefficiency of
photodetectors does not affect the quality of the generated
entangled states, but it decreases the success probability. In our
scheme the success probability of the state to obtain the GESs
$|\overline{{\chi}^{'}}\rangle$
 and $|\overline{{\chi}^{''}}\rangle$ are $\eta^{2}$ when the quantum efficiency of each photodetector is $\eta$.

\section{Concluding remarks}
In conclusion, we have proposed a theoretical scheme to generate
genuine entangled states of four atomic qubits in separated
optical cavities using the atom-light interaction under the
condition of the large detuning and single photon detections. We
have shown that GESs of four atomic qubits can be produced
deterministically. Starting with one prepared GES we have found
the sixteen orthonormal and independent GESs. We have shown that
these sixteen GESs build a new type of representation the
four-qubit system, the genuine entangled-state representation.
This representation provides us with new interesting insight into
better understanding multipartite entanglement. It have indicated
that the GHZ state and $W$ state, the cluster state, and the
symmetric Dicke state for a four-qubit system can be explicitly
expressed in terms of the sixteen GESs. We have considered the
influence of the imperfection of photodetectors in the present
scheme, and indicated that  the inefficiency of photodetectors
does not affect the quality of the generated entangled states, but
it decreases the success probability. It is believed that the GESs
created in the present scheme provide new entanglement sources to
realize quantum information processing.

\acknowledgments This work was supported by the National
Fundamental Research Program Grant No.  2007CB925204, the National
Natural Science Foundation under Grant Nos. 10775048 and 10325523,
and the Education Committee of Hunan Province under Grant No.
08W012.

\end{document}